\documentclass[prl,aps,twocolumn,superscriptaddress,floats,showpacs,final, fleqn,hidelinks]{revtex4-2}
\usepackage{dcolumn}
\usepackage{bm}
\usepackage{color}
\usepackage{mathtools}
\usepackage{amsmath,amsfonts,amssymb,bm}
\usepackage{graphicx,tikz}
\usepackage[pdftex, bookmarks=true, linktoc=section, pdfstartview=FitH]{hyperref} 

\newcommand{\be}{\begin{equation}}
\newcommand{\ee}{\end{equation}}
\newcommand{\ba}{\begin{eqnarray}}
\newcommand{\ea}{\end{eqnarray}}

\begin{document}

\title{Transient Dynamical Phase Diagram of the Spin--Boson model}

\author{Olga Goulko}
\email{olga.goulko@umb.edu}
\affiliation{Department of Physics, University of Massachusetts Boston, Boston Massachusetts 02125, USA}
\author{Hsing-Ta Chen}%
 \email{hchen25@nd.edu}
\affiliation{%
Department of Chemistry and Biochemistry, University of Notre Dame, Notre Dame, Indiana 46556, USA
}%

\author{Moshe Goldstein}
    \email{mgoldstein@tauex.tau.ac.il }
    \affiliation{
	School of Physics, Tel Aviv University, Tel Aviv 6997801, Israel
    }    

\author{Guy Cohen}
    \email{gcohen@tau.ac.il}
    \affiliation{
	School of Chemistry, Tel Aviv University, Tel Aviv 6997801, Israel
    }

\date{\today}

\begin{abstract}
We investigate the real-time dynamics of the sub-Ohmic spin--boson model across a broad range of coupling strengths, using the numerically exact inchworm quantum Monte Carlo algorithm.
From short- and intermediate-time dynamics starting from an initially decoupled state, we extract signatures of the zero-temperature quantum phase transition between localized and delocalized states.
We show that the dynamical phase diagram thus obtained differs from the equilibrium phase diagram in both the values of critical couplings and the associated values of the critical exponent.
We also identify and quantitatively analyze two competing mechanisms for the crossover between coherent oscillations and incoherent decay. Deep in the sub-Ohmic regime, the crossover is driven by the damping of the oscillation amplitude, while closer to the Ohmic regime the oscillation frequency itself drops sharply to zero at large coupling.
\end{abstract}

\maketitle

\paragraph{Introduction.}
The spin--boson model describes a two-level system, or spin, coupled to a continuum of bosonic modes.
It is foundational in our understanding of quantum phase transitions (QPT) \cite{leggettreview}, embodying the standard framework for studying environmental dissipation \cite{quantumdissbook} in chemical dynamics \cite{nitzan_chemical_2006}, quantum optics \cite{de_vega_dynamics_2017}, and quantum information science \cite{breuer_colloquium_2016}.
The effect of the bosonic environment on the system is often described by the spectral density $J(\omega)\propto\omega^s$ for frequencies below a certain cutoff $\omega_c$.
The physically rich sub-Ohmic regime, $0<s<1$, can be realized in various physical systems, including superconducting and mesoscopic circuits \cite{tong2006mesoscopicring,yu2012supercondcircuits,drivenSBexp2017, leppakangas2018supercondircuit, yamamoto2019supercondcircuits} and trapped ion systems \cite{porras2008mesoscopic, lemmer2018trapped}.

Analytical solutions of the spin--boson model exist in only a few special cases.
It is mostly accepted that in the $\omega_c\rightarrow\infty$ limit, the sub-Ohmic system is always localized at zero temperature, and always delocalized at finite temperature \cite{leggettreview}.
The behavior at finite $\omega_c$ has received much attention, in particular due to the existence of a QPT between the localized and the delocalized phases.
One expects the QPT of the spin--boson model to be in the same universality class as the thermal phase transition of the classical Ising spin chain with long-ranged interactions \cite{quantumdissbook,SBIsing1972,SBIsing1997,SBtoIsingCorresp}.
After some initial controversy about the validity of the correspondence \cite{BullaEtAlCritExp2003,VojtaTongBulla2005, VojtaTongBullaErratum2009, Vojta2012NRGerrors} critical exponents have been confirmed to match the Ising model prediction \cite{criticalExpSB,AlvermannFehske2009,GuoWeichselbaumDelftVojta2012,shen2023variational}.  

Where analytical solutions are unavailable, a wide array of numerical methods have been employed to simulate the dynamics of the spin--boson model, including the multilayer multiconfiguration time-dependent Hartree (ML-MCTDH) theory \cite{wang2003,wang2008}, the quasi-adiabatic propagator path integral (QuAPI) method \cite{Makri1992,Makri1993,Makri2017,Makri2023}, and the hierarchy equation of motion (HEOM) approach \cite{Tanimura1989,Tanimura2005,YiJing2011,Qiang2014,Cao2013, HEOM2017}.
Interest within quantum information science has also resulted in several successful approaches \cite{strathearn_efficient_2018,popovic_quantum_2021,cygorek_simulation_2022,gribben_exact_2022}.
Focusing on the sub-Ohmic regime, we employ the numerically exact inchworm Quantum Monte Carlo (QMC) method \cite{inchworm, thetainchworm1, thetainchworm2,cai_numerical_2020,kim_pseudoparticle_2022,cai_inchworm_2023} in this work.
This provides controlled results in a wide range of sub-Ohmic exponents, coupling strengths, frequency cutoffs, and temperatures.

Although at finite $\omega_c$ delocalization is possible even at zero temperature, it can take a very long time for the system to delocalize \cite{leggettreview}.
Studies of the QPT have thus so far been performed directly in equilibrium \cite{BullaEtAlCritExp2003,VojtaTongBulla2005, VojtaTongBullaErratum2009, Vojta2012NRGerrors,criticalExpSB,AlvermannFehske2009,GuoWeichselbaumDelftVojta2012,shen2023variational} or in the infinite time limit \cite{HEOM2017}, even though experimental probing of the dynamics can only access short- and intermediate-time properties.
Another important feature manifested by the transient dynamics is the coherence of the decay process \cite{KastAnkerhold2013,omegacDep2013,HEOM2017,otterpohl2022sb,Lipeng2023}, which is always probed in terms of short-time dynamics. Usually, the change from well-defined oscillations at weak spin--bath coupling to incoherent decay at strong coupling is characterized as a crossover \cite{KastAnkerhold2013,omegacDep2013,otterpohl2022sb,Lipeng2023}.

In this Letter, we extract signatures of the localization transition directly from the dynamics, showing that the result differs dramatically from the localization transition in equilibrium for $s\gtrsim0.4$.
We also reveal two distinct mechanisms that drive the change from coherent to incoherent decay: Overdamping of the oscillation amplitude that occurs at all values $0<s<1$; and a sharp decrease in oscillation frequency that is only observed for $s\gtrsim0.5$.
While the former is indeed a smooth crossover, the latter mechanism has the hallmarks of a sudden transition.

\paragraph{Model.}
Setting $\hbar=1$, the spin--boson model is described by the Hamiltonian $H=H_\mathrm{s}+H_\mathrm{b}+H_\mathrm{sb}$, where
\begin{equation}
\begin{aligned}
H_{\mathrm{s}} & =\frac{\epsilon}{2}\hat{\sigma}_{z}+\frac{\Delta}{2}\hat{\sigma}_{x},\\
H_{\mathrm{b}} & =\sum_{\ell}\omega_{\ell}\left(b_{\ell}^{\dagger}b_{\ell}+\frac{1}{2}\right),
\end{aligned}
\label{eq:system-bath-hamiltonian}
\end{equation}
are the system and bath Hamiltonians, respectively; while
\begin{equation}
\begin{aligned}H_{\mathrm{sb}} & =\frac{\hat{\sigma}_{z}}{2}\sum_{\ell}c_{\ell}x_{\ell}=\frac{\hat{\sigma}_{z}}{2}\sum_{\ell}\frac{c_{\ell}}{\sqrt{2\omega_{\ell}}}\left(b_{\ell}^{\dagger}+b_{\ell}\right)\end{aligned}
\label{eq:hybridization-hamiltonian}
\end{equation}
describes the system--bath coupling or hybridization between them, which is linear in the bath coordinates, $x_\ell$.
$\hat{\sigma}_i$ are Pauli matrices, $\epsilon$ is the bias, $\Delta$ is the tunneling amplitude, and $b_\ell$ ($b^\dagger_\ell$) are the bosonic annihilation (creation) operators.
The coupling constants $c_\ell$ control the interaction strength between the spin and the harmonic mode of frequency $\omega_\ell$.
In the bath thermodynamic limit, one can characterize the system--bath coupling by defining the continuous spectral density 
\be
J\left(\omega\right)=\frac{\pi}{2}\sum_{\ell}\frac{c_{\ell}^{2}}{\omega_{\ell}}\delta\left(\omega-\omega_{\ell}\right)=2\pi\alpha\omega^s\omega_c^{1-s}e^{-\omega/\omega_c},\label{eq:spectral}
\ee
where $\alpha$ is the coupling strength, $\omega_c$ the cutoff frequency, and $0<s<1$ is the sub-Ohmic exponent.
The cutoff frequency is associated with the relaxation timescale of the harmonic bath, $2\pi/\omega_c$.

The transient behavior of the sub-Ohmic spin--boson model depends strongly on the bath initial condition \cite{KastAnkerhold2013}.
One usually considers the ``decoupled'' initial condition (where the bath is decoupled from the spin subsystem) or the ``shifted'' initial condition (where the bath is at equilibrium with the spin subsystem state fixed) \cite{Lipeng2023}.
Here, the initial system is assumed to be decoupled and the total density matrix can be represented by the factorized state $\rho_0=\rho_\mathrm{s}\otimes\rho_\mathrm{b}$, where the spin subsystem is in state $1$ ($\rho_\mathrm{s}=|1\rangle\langle1|$) and the bath in thermal equilibrium, $\rho_\mathrm{b}=e^{-\beta H_\mathrm{b} }/\textnormal{Tr}\left\{e^{-\beta H_\mathrm{b}}\right\}$.
Following the initial preparation, we focus on the dynamics of the population difference between the two spin states, $\langle\hat{\sigma}_z(t)\rangle=\textnormal{Tr}\left\{\rho_0e^{iHt}\hat{\sigma}_ze^{-iHt}\right\}$.

\paragraph{Method.}
To obtain numerically exact dynamics of $\langle\hat{\sigma}_z(t)\rangle$, we employ a real-time variant \cite{muhlbacher_real-time_2008, schiro_real-time_2009, werner_diagrammatic_2009, schiro_real-time_2010, gull_numerically_2011, cohen_memory_2011, cohen_numerically_2013, cohen_greens_2014-1, cohen_greens_2014, vanhoecke_diagrammatic_2023,muhlbacher_real-time_2008, schiro_real-time_2009, werner_diagrammatic_2009, schiro_real-time_2010, gull_numerically_2011, cohen_memory_2011, cohen_numerically_2013, cohen_greens_2014-1, cohen_greens_2014, vanhoecke_diagrammatic_2023} of the continuous-time QMC algorithm \cite{rubtsov_continuous-time_2004,werner_continuous-time_2006,gull_continuous-time_2011}.
To bypass the dynamical sign problem, which makes it exponentially difficult to access long times, we implemented an ``inchworm" algorithm \cite{inchworm}.
The idea behind this algorithm is that short-time propagators are less expensive to calculate than long-time propagators, and can be recycled to construct easier expansions for propagators over longer timescales in subsequent Monte Carlo steps.
The inchworm algorithm has been successfully applied to study the dynamics of the spin--boson model with the Debye spectral density \cite{thetainchworm1,thetainchworm2}, in which two types of diagrammatic expansions were developed: the spin--bath coupling expansion and the diabatic coupling expansion (the latter is combined with a cumulant inchworm expansion).
The results are consistent, although in different parameter regimes one of the expansions may be more efficient than the other \cite{thetainchworm2}.
In the context of the Ohmic and sub-Ohmic spin--boson model, we validated the accuracy of our inchworm Monte Carlo results by detailed comparisons with numerically exact results that are available at zero temperature using ML-MCTDH \cite{wang2010coherent}, see Supplemental Material (SM) \cite{supp} for more details.

\begin{figure}[t]
    \centering
    \includegraphics[width=\columnwidth]{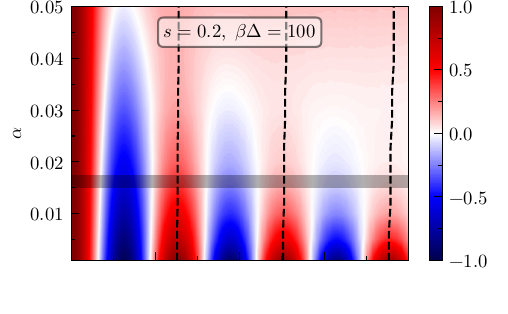}\vspace*{-8mm}
    \includegraphics[width=\columnwidth]{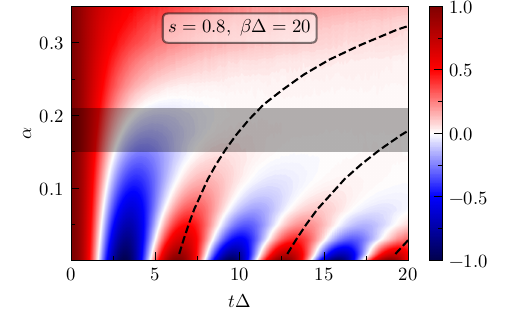}
    \caption{
    Time-evolution of $\langle\sigma_z(t)\rangle$ as a function of coupling $\alpha$ deep in the sub-Ohmic regime (top panel) and close to the Ohmic regime (bottom panel).
    The horizontal shaded regions indicate the approximate location of the localization/delocalization transition, $\alpha^*$, extracted from dynamical data.
    Two competing mechanisms for the coherence/incoherence crossover are visible: the damping of the oscillation amplitude (both panels), and the decrease in oscillation frequency (bottom panel only), see text for discussion.
    Dashed black lines denote the analytical prediction for peaks in the dynamics, based on \cite{kehrein1996} (see SM \cite{supp}).
    }
    \label{fig:dynamics}
\end{figure}
To extract quantitative observables, we fit the dynamics of $\langle\sigma_z(t)\rangle$ to the following functional form,
    \be
    \langle\sigma_z(t)\rangle \approx a\cos(\Omega t+\phi) e^{-\gamma_1 t} + be^{-\gamma_2t} + c.\label{eq:fit}
    \ee
This heuristic choice of fitting function is well-suited to capture several key characteristics: 
(i) The first term describes the damped oscillation at the renormalized frequency $\Omega$ and the damping coefficient $\gamma_1$; 
(ii) The second term describes the overall decay at the decay rate $\gamma_2$;
(iii) The long-time behavior is captured by the offset coefficient $c=\langle\sigma_z(t\rightarrow\infty)\rangle$.
Given the fitting coefficients, we can identify different regimes by using the following criteria:
(a) The long-time population is considered \emph{localized} if $c\neq0$ and \emph{delocalized} if $c=0$. The localization/delocalization transition can be delineated as the boundary line of the $c=0$ region.
(b) The damped oscillation becomes \emph{incoherent} if $\Omega/\Delta\rightarrow0$. In this case, the coherence/incoherence crossover is driven by the renormalized frequency decreasing, which corresponds to the boundary of the $\Omega/\Delta=0$ region.
(c) Alternatively, the transient dynamics is also regarded \emph{incoherent} when the oscillation is over-damped, i.e.\ $\Omega/\gamma_1<1$. In this case, the coherence/incoherence crossover is driven by the damping, which corresponds to the $\Omega/\gamma_1=1$ line in the parameter space.

\paragraph{Results.}
We only consider unbiased systems, i.e., $\epsilon=0$.
We set $\omega_c=10\Delta$ throughout and use $\Delta$ as the unit of frequency.
We have found the following inverse temperatures to be indistinguishable from the zero temperature limit within the statistical errors of the simulation: $\beta\Delta=100$ for $0.2\leq s<0.5$, and $\beta\Delta=20$ for $0.5\leq s\leq 1$.

In Fig.~\ref{fig:dynamics} we qualitatively compare the time-evolution of $\langle\sigma_z(t)\rangle$ as a function of coupling $\alpha$ in the deep sub-Ohmic ($s=0.2$) and near-Ohmic ($s=0.8$) regimes.
In both regimes, the system undergoes a transition from a delocalized state at weak coupling to a localized state at strong coupling.
The corresponding critical coupling $\alpha^*$, as obtained by criterion (a), increases with increasing $s$. 
We also observe the crossover between coherent oscillations at weak coupling and incoherent decay at strong coupling.
In the deep sub-Ohmic regime, the crossover is damping-driven as the oscillation frequency remains essentially unchanged and the amplitude is more strongly damped when $\alpha$ increases (criterion (c)).
In the near-Ohmic regime on the other hand, in addition to the decrease in amplitude, the oscillation frequency also decreases with increasing $\alpha$ and completely vanishes at large $\alpha$ (criterion (b)).
The $\alpha$ dependence of the renormalized oscillation frequency can be captured qualitatively by a theoretical treatment based on the analytical renormalization group method used in Ref.~\cite{kehrein1996}; the result is displayed as the dashed black curves representing oscillation peaks in Fig.~\ref{fig:dynamics} (see SM~\cite{supp} for further discussion).

\begin{figure}
    \centering
    \includegraphics[width=\columnwidth]{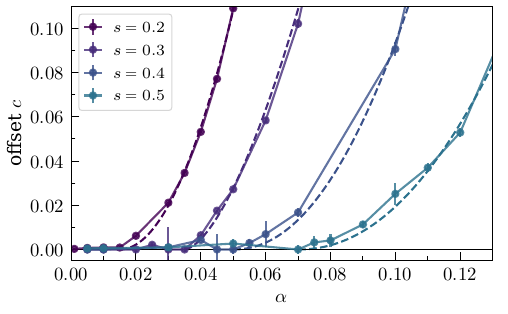}
    \caption{
        The offset fit coefficient $c$ as a function of the coupling $\alpha$ for different values of $s\leq0.5$.
        A sharp transition between the localized and delocalized regimes is visible, but the associated scaling does not correspond to the critical exponent of the equilibrium QPT. The dashed lines show corresponding power law fits, with fit exponents ranging between 1.6 and 2.1.
        }
    \label{fig:offset}
\end{figure}

A quantitative analysis of the fit parameters obtained by fitting the numerical data to Eq.~\eqref{eq:fit} sheds light on the nature of the localization/delocalization transition and the coherence/incoherence crossover.
To distinguish between the localized and delocalized phase we examine the offset coefficient $c$ of the fit, shown in Fig.~\ref{fig:offset}.
A sharp transition from zero to finite $c$ is observed for $s\lesssim0.5$.
For larger $s$ (shown in Fig.~S5 of the SM~\cite{supp}), the offset $c$ increases more smoothly with $\alpha$, resulting in a larger uncertainty on the critical value $\alpha^*$.
For the equilibrium QPT the critical exponents of the transition are known \cite{quantumdissbook,SBIsing1972,SBIsing1997,SBtoIsingCorresp}.
At $s\leq0.5$ the exponent $\beta$ defined via $\langle\sigma_z\rangle\propto(\alpha-\alpha_c)^\beta$ equals $1/2$, implying that the offset is a concave function of $\alpha$ \cite{GuoWeichselbaumDelftVojta2012}.
In contrast, the corresponding scaling of the transient phase diagram analyzed in this work indicates a scaling exponent above 1, as the offset is a convex function of $\alpha$, see Fig.~\ref{fig:offset} and SM~\cite{supp} for more information.

\begin{figure}
    \centering
    \includegraphics[width=\columnwidth]{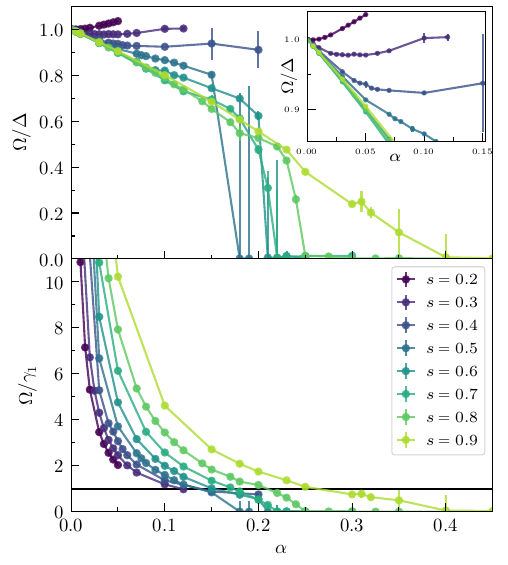}
    \caption{
        Two distinct hallmarks of the coherence/incoherence crossover: the renormalized oscillation frequency, $\Omega/\Delta$, (top panel) and the ratio between $\Omega$ and the corresponding damping coefficient $\gamma_1$ (bottom panel).
        The inset in the top panel magnifies the region where the frequency behaves non-monotonically for small values of $s$. The horizontal line in the bottom panel marks the $\Omega/\gamma_1=1$ threshold.
        }
    \label{fig:incoherence}
\end{figure}

Next, to characterize the coherence/incoherence crossover, we focus on the $\alpha$-dependence of the renormalized frequency $\Omega/\Delta$, criterion (b), and its ratio to the damping coefficient $\Omega/\gamma_1$, criterion (c), both shown in Fig.~\ref{fig:incoherence}. 
Interestingly, after a small initial dip, the frequency increases with increasing $\alpha$ for small values of $s$ ($s<0.5$).
In contrast, for $s\gtrsim0.5$ the frequency monotonically decreases with $\alpha$ and eventually sharply drops to zero at sufficiently strong coupling. This sharp frequency drop is not predicted by the renormalization group treatment \cite{kehrein1996,supp} and is distinct from the localization/delocalization transition.
The widely accepted notion is that the crossover between coherent and incoherent dynamics is smooth. As opposed to the sharp frequency-driven incoherence transition, criterion (c) identifies the standard hallmark of a smooth incoherence crossover, which is driven by the over-damping of the oscillation. The $\Omega/\gamma_1$ ratio decreases with increasing $\alpha$ for all values of $s$ and we choose the (arbitrary) threshold $\Omega/\gamma_1=1$ to indicate the over-damping crossover. The two incoherence mechanisms are clearly distinct from each other, both in terms of their sharpness and in terms of the parameter range in which they were observed to occur.

\begin{figure}
    \centering
    \includegraphics[width=\columnwidth]{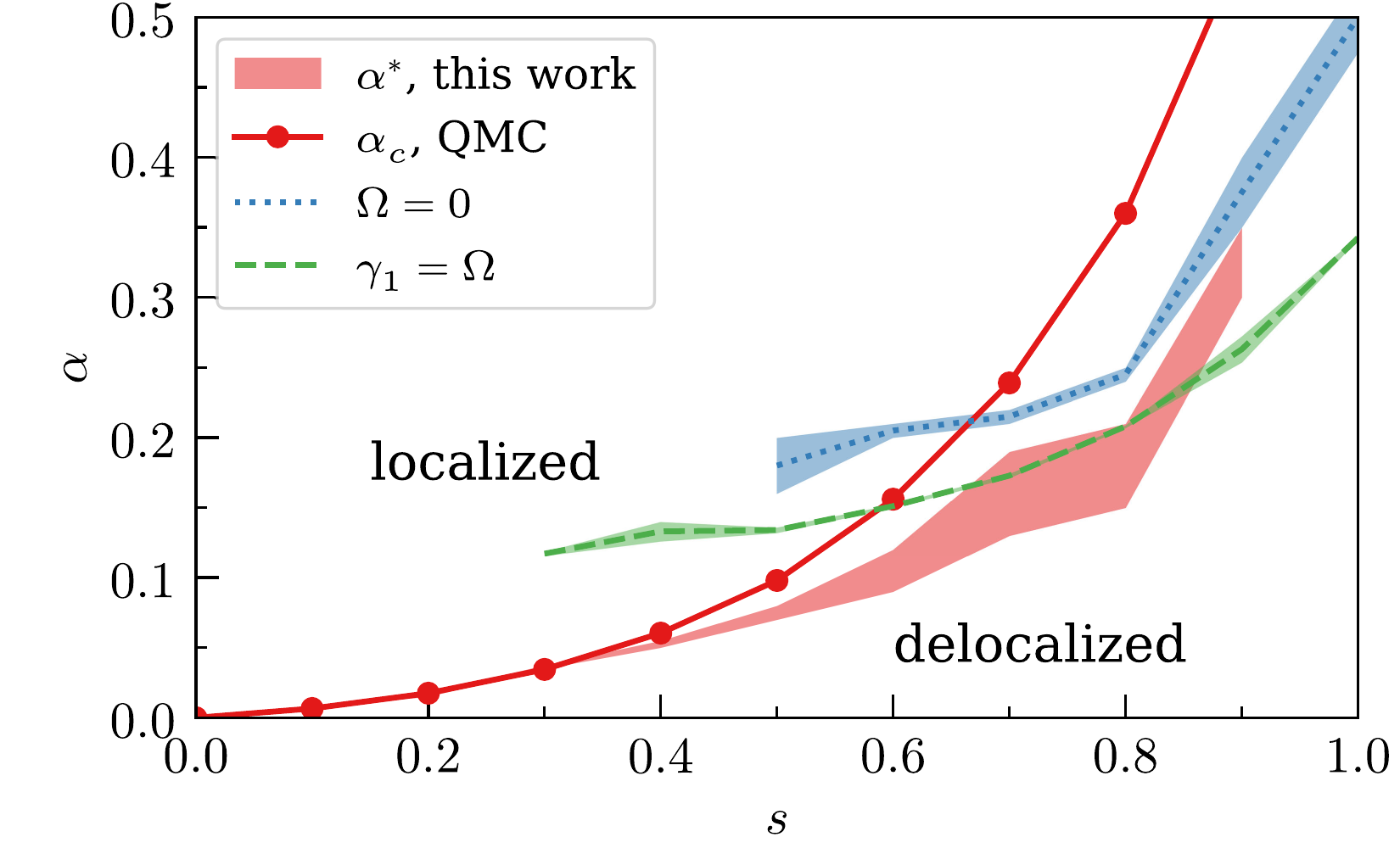}\hfill
    \caption{
    Transient dynamical phase diagram of the sub-Ohmic spin--boson model.
    The red shaded region is our numerical result for $\alpha^*$ that characterizes the dynamical localization-delocalization transition (criterion (a)), with the area marking a confidence interval.
    For $s\lesssim0.4$ it agrees well with the equilibrium $\alpha_c$ from a QMC calculation for a model with sudden cutoff \cite{criticalExpSB} (red circles and solid line) \cite{footnote}.
    The dotted and dashed lines are the two characterizations of the change from coherent to incoherent dynamics, corresponding to criteria (b) and (c), respectively.
    The system tends toward incoherence above the lines, i.e.\ at higher values of $\alpha$.
    The frequency-driven crossover (blue dotted line) was only observed in the near-Ohmic regime, for the range of parameters considered.
    The damping-driven crossover (green dashed line), which we defined as $\gamma_1=\Omega$, was observed in a wider parameter range.
    }
    \label{fig:phasediag}
\end{figure}

A summary of our quantitative results is presented through the phase diagram in the parameter space of $\alpha$ and $s$, see Fig.~\ref{fig:phasediag}.
The transient critical coupling $\alpha^*$ distinguishing the localized and delocalized phases agrees well with the $\alpha_c$ of the equilibrium QPT for a model with sudden frequency cutoff \cite{Bruognolo2013masters, criticalExpSB, wong2008density} for sufficiently small $s$, where the critical couplings are also well-described by the analytical prediction from the generalized polaron ansatz \cite{entanglementSB2011}.
For $s\gtrsim0.4$ the two values begin to deviate.
The frequency- and the damping-driven crossover lines depicted in Fig.~\ref{fig:phasediag} are clearly distinct from each other, with the frequency-driven crossover only observed to occur in the near-Ohmic regime, while the damping-driven crossover can occur at any value of $s$ (though it is numerically challenging to observe at small $s$).
Coherent and incoherent regimes are observed on either side of the localization/delocalization line. For the Ohmic spin--boson model the location of the incoherence transition is known analytically to occur at $\alpha=0.5$ (Toulouse point) \cite{quantumdissbook}. Our numerical results for the frequency-driven transition (criterion (b)) are consistent with the Toulouse point. The $\Omega=\gamma_1$ line of the damping-driven criterion (c) yields a different value for the Ohmic transition point.

\paragraph{Conclusions.}
We extracted the transient dynamical phase diagram of the sub-Ohmic spin--boson model from numerically exact data for $\langle\sigma_z(t)\rangle$ at short and intermediate times, which are the experimentally relevant regimes.
Similarly to the corresponding equilibrium phase diagram, which has been extensively studied, the system features a transition between localized and delocalized regimes.
The corresponding critical couplings agree well with the equilibrium values for $s\lesssim0.4$, but deviate for larger values of $s$.
The equilibrium critical exponents, on the other hand, are not reproduced from the intermediate-time data: rather, we find apparent critical behavior with different exponents.
We also studied the change from coherent to incoherent decay, typically characterized as a crossover in the literature.
We identified two distinct mechanisms driving this change: the reduction of the oscillation frequency itself (at $s\gtrsim0.5$) and the damping of the oscillation amplitude (at all values of $s$).
While the latter is smooth, the drop in frequency occurs sharply and is thus more reminiscent of a phase transition than a crossover.

The inchworm algorithm used here is efficient over a wide range of parameters.
Looking forward, it can also be used to explore dynamics and full charge/energy counting statistics at higher temperatures \cite{ridley_numerically_2018,ridley_numerically_2019,ridley_lead_2019,erpenbeck_revealing_2021}, in more general models \cite{eidelstein_multiorbital_2020}, or in the presence of intrinsically nonequilibrium drives such as multiple baths at different thermodynamic parameters.
Extensions exist for investigating dynamics over very long timescales \cite{pollock_reduced_2022} and nonequilibrium steady states \cite{erpenbeck_quantum_2023,erpenbeck_shaping_2023}.
On a more general note, our work points the way toward the ability to investigate and potentially control transient ``phase diagrams'' in numerous scenarios, opening up intriguing prospects for revealing new physics in experiments, theory and simulations.

\paragraph{Acknowledgments. }
We thank Jianshu Cao, Jan von Delft, Carlos Gonz\'alez-Guti\'errez, Olivier Parcollet, Nikolay Prokof'ev, and Andreas Weichselbaum for inspiring discussions and the authors of Refs.~\cite{criticalExpSB} and \cite{wang2010coherent} for sharing their data.
O.G. is supported by the NSF under Grants No.~PHY-2112738 and No.~PHY-2328774.
M.G. is supported by the Israel Science
Foundation (ISF) and the Directorate for Defense Research
and Development (DDR\&D) Grant No. 3427/21, by the ISF grant No. 1113/23
and by the U.S.-Israel Binational Science Foundation
(BSF) Grant No. 2020072.
G.C. is supported by the ISF (Grant No.~2902/21) and by the PAZY foundation (Grant No.~318/78).
We acknowledge high-performance computing support of the R2 compute cluster (DOI: 10.18122/B2S41H) provided by Boise State University's Research Computing Department, the chimera cluster provided by UMass Boston Research Computing, and the Unity cluster.

\bibliography{spinboson}

\begin{thebibliography}{77}
\bibitem [{foo()}]{footnote}%
  \BibitemOpen
  \href@noop {} {}\bibinfo {note} {Similar results for this model were also
  obtained with a Variational Matrix Product State (VMPS) calculation
  \cite{Bruognolo2013masters}.}\BibitemShut {Stop}%
\end{thebibliography}

\clearpage
\onecolumngrid
\appendix
\newpage
\begin{center}\textbf{Supplemental Material to ``Transient dynamical phase diagram of the spin--boson model''}\end{center}

\setcounter{equation}{0}
\setcounter{figure}{0}
\setcounter{page}{1}
\makeatletter
\renewcommand{\theequation}{S\arabic{equation}}
\renewcommand{\thefigure}{S\arabic{figure}}

\section{Numerical benchmarks}
We have performed an extensive array of tests to ensure the validity of our numerical results across the range of physical parameters considered.

For internal consistency, we employed two independent diagrammatic expansions: the spin–bath coupling expansion and the diabatic coupling expansion combined with a cumulant inchworm expansion.
We also varied all auxiliary numerical parameters, such as time step and maximal expansion order, to exclude potential truncation errors. For the spin-bath coupling expansion a maximal order of 4 was found to be sufficient to achieve convergence and to produce accurate results. For the cumulant expansion the maximal order was set to 14, but in practice, such high orders were never generated by the simulations, so that effectively the maximal order can be considered to be infinite.

The maximal times simulated varied from $t_{max}\Delta=20$ to $t_{max}\Delta=80$, depending on the dynamical behavior of the system.  Larger values of $t_{max}$ were needed in strongly incoherent regimes that delocalize slowly (large $s$ and large $\alpha$), as well as in strongly coherent regimes with small damping (small $s$ and small $\alpha$). We checked consistency of our fit results for different values of $t_{max}$ for representative physical parameter sets.

In the Ohmic regime, we reproduce the analytically known Toulouse point and ensure consistency with numerical data from previous work with the inchworm algorithm \cite{thetainchworm1, thetainchworm2}, which was in turn benchmarked with QUAPI and HEOM.

\begin{figure*}[b]
    \centering
    \includegraphics[width=0.5\textwidth]{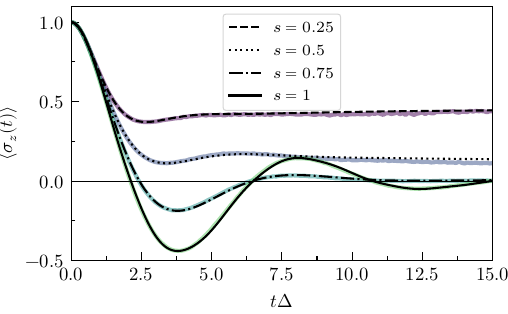}\hfill
    \includegraphics[width=0.5\textwidth]{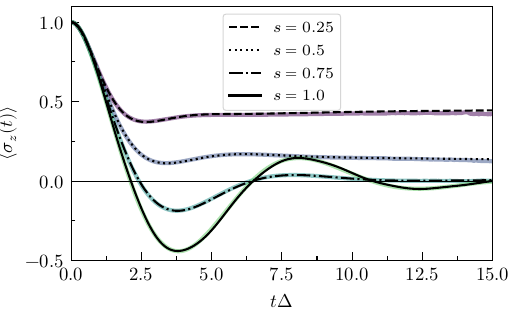}
    \caption{
    Time-evolution of $\sigma_z(t)$ for $\omega_c=5\Delta$ and $\alpha=0.2$ for different values of $s$.
    The black lines correspond to data from Fig.~10(a) of \cite{wang2010coherent} (note that units were adjusted to match our convention for the model Hamiltonian).
    They agree well with the inchworm QMC data from this work (color-coded) for both the spin-bath coupling expansion (left panel) and diabatic cumulant expansion (right panel).}
    \label{fig:wangthoss}
\end{figure*}

Figure~\ref{fig:wangthoss} compares our data at very low temperature ($\beta\Delta=100$) to the numerically exact results obtained at zero temperature using ML-MCTDH \cite{wang2010coherent}.
Our dynamics data also agree with the data for the decoupled initial condition shown in Fig.~5 of Ref.~\cite{KastAnkerhold2013}.

\section{Comparison with analytical estimates at low coupling}
\begin{figure}
    \centering
    \includegraphics[width=0.6\textwidth]{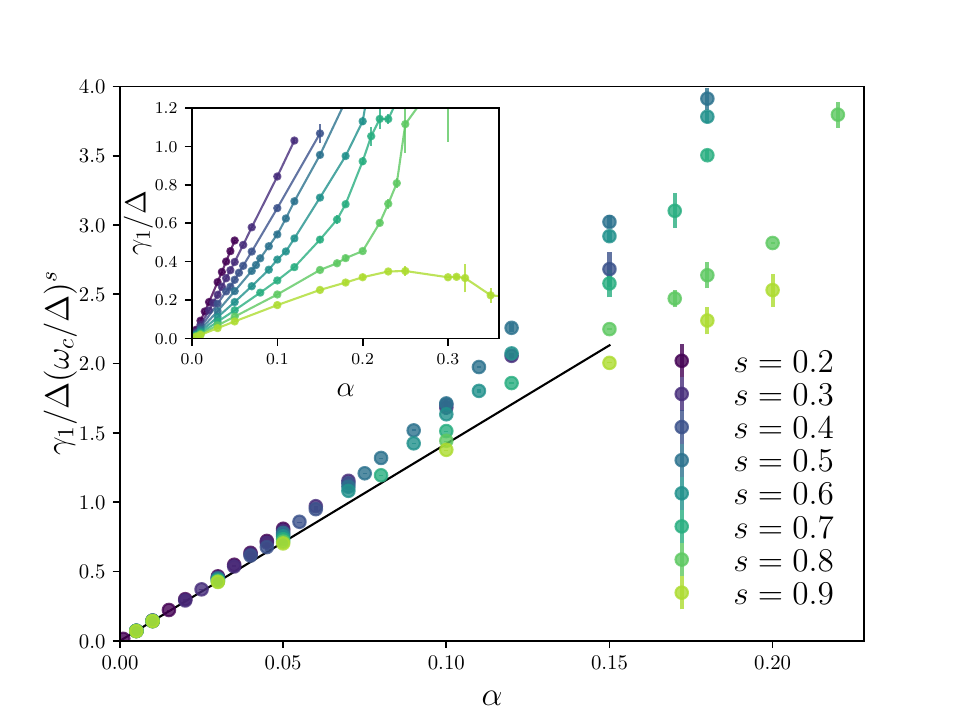}
    \caption{The damping coefficient $\gamma_1$ of the coherent part of the time evolution as function of $\alpha$ for different values of $s$. For small $\alpha$ the damping exhibits the predicted scaling $\gamma_1\propto\alpha(\Delta/\omega_c)^s$.}
    \label{fig:gamma1}
\end{figure}

In the context of the Redfield equation (based on Born-Markov approximation), the damping coefficient of the oscillating dynamics can be estimated by Fermi's golden rule (FGR) \cite{nitzan_chemical_2006}
\ba
    \gamma_{FGR}=\frac{1}{4}\int_{-\infty}^\infty dt e^{i\Delta t}C_b(t).
\ea
Here we write the system-bath coupling in terms of $H_\mathrm{sb}=\frac{1}{2}\hat{\sigma}_{z}\sum_{\ell}c_{\ell}x_{\ell}\equiv\hat{V}_sV_b$, where the system operator is $\hat{V}_s=\frac{1}{2}\hat{\sigma}_{z}$ and the bath operator $V_{b}=\sum_{\ell}c_{\ell}x_{\ell}$.
In the absence of the system, the autocorrelation function of the bath operator $C_b(t)=\text{Tr}\{\rho_b V_b(t)V_b(0)\}$ is given by
\ba
C_b(t)=\frac{1}{2\pi}\int_0^\infty d{\omega}J(\omega)\left[\coth(\frac{\beta\omega}{2})\cos\omega{t}-i\sin\omega{t}\right].
\ea
In the low-temperature limit ($\coth(\frac{\beta\omega}{2})\approx1$), we can carry out the time integration first $\int_{-\infty}^\infty e^{i\Delta t}e^{-i\omega t}=2\pi\delta(\Delta-\omega)$, which results in the spectral function evaluated at $\omega=\Delta$.
Given the spectral density $J(\omega)$ in Eq.~\eqref{eq:spectral}, the coherence decay rate can be estimated by
\ba
\gamma_{FGR}=\frac{\pi}{2}\alpha\Delta^{s}\omega_c^{1-s}e^{-\Delta/\omega_c}\propto\alpha\left(\Delta/\omega_c\right)^s,
\ea
\ba
\frac{\gamma_{FGR}}{\Delta}\left(\frac{\omega_c}{\Delta}\right)^s=\frac{\pi}{2}\alpha\frac{\omega_c}{\Delta}e^{-\Delta/\omega_c}\approx 14.21\alpha,
\ea
which agrees with the scaling in Fig.~\ref{fig:gamma1} when $\alpha$ is small. Note that the FGR decay rate is only valid for small $\alpha$.

\begin{figure}
    \centering
    \includegraphics[width=0.5\textwidth]{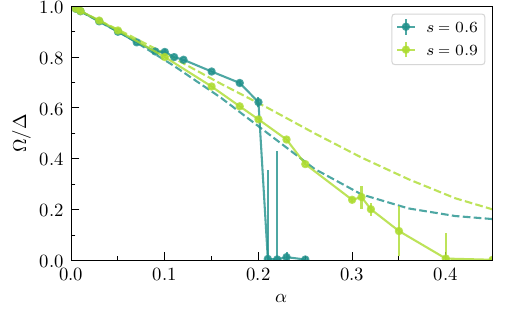}\hfill
    \includegraphics[width=0.5\textwidth]{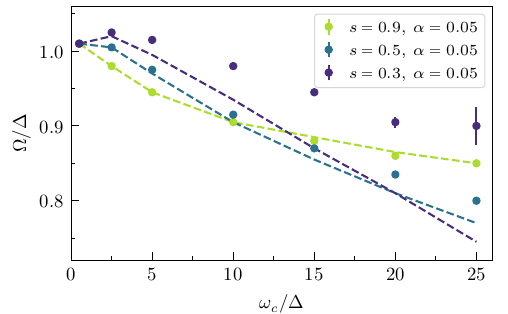}
    \caption{Left panel: Oscillation frequency $\Omega$ as function of coupling $\alpha$ for two values of $s$ in the regime that exhibits the frequency-driven coherence/incoherence transition. The symbols represent numerical data (solid lines are guides to the eye) and the dashed lines of the corresponding color are analytical predictions from \cite{kehrein1996}. The approximate theory and numerics agree at weak coupling. The numerically observed sharp drop to zero frequency at the transition is not predicted by theory.
    Right panel: Oscillation frequency $\Omega$ as function of the cut-off frequency $\omega_c$ at fixed $\alpha$ and for different values of $s$. The symbols represent numerical data and the dashed lines of the corresponding color are predictions based on numerically solving Eq.~(11) from~\cite{kehrein1996}. The agreement between the approximate theory and numerics is especially good closer to the Ohmic regime, which corresponds to weaker coupling when $\alpha$ is held fixed.}
    \label{fig:kehrein}
\end{figure}

Figure~\ref{fig:kehrein} compares the numerically found oscillation frequency $\Omega$ with the analytical prediction from \cite{kehrein1996}. For a direct comparison, the analytical predictions were obtained by numerically solving Eq.~(11) from \cite{kehrein1996} with our form of spectral density $J(\omega)$ and for the specific parameter values used in the simulation. Our results agree with the analytical estimate at sufficiently weak coupling, but not at strong coupling. In particular, the setup from \cite{kehrein1996} predicts a smeared-out crossover for all $s<1$, rather than the sharp drop of $\Omega$ that was numerically found at large coupling.  We also compare our numerical data with the analytical prediction from \cite{kehrein1996} as a function of cut-off frequency $\omega_c$ for $\alpha=0.05$ and different values of $s$. The agreement is best at small values of $\omega_c$ and in the near-Ohmic regime, since $\alpha=0.05$ corresponds to weak coupling for large values of $s$, but to strong coupling for small values of $s$.

\section{Details on the fitting procedure}
\begin{figure*}
    \centering
\includegraphics[width=0.5\textwidth]{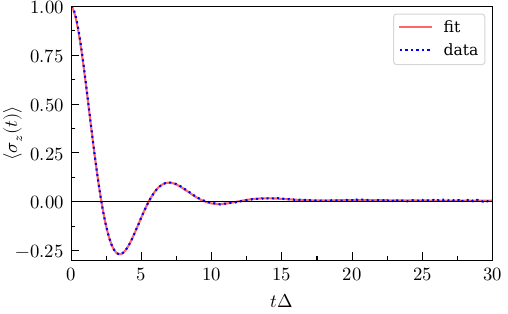}\hfill
\includegraphics[width=0.5\textwidth]{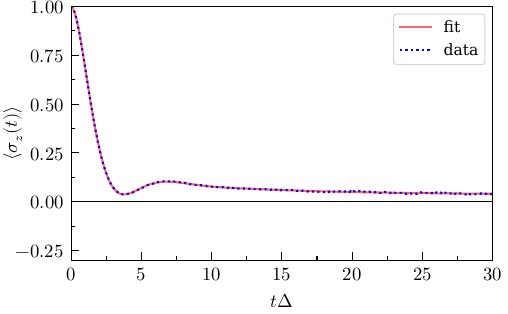}
    \caption{Examples of fits to the simulation data $\langle\sigma_z(t)\rangle$ for $s=0.5$, $\beta\Delta=20$, $\omega_c=10\Delta$. Left panel: $\alpha=0.07$ (coherent and delocalized); Right panel: $\alpha=0.11$ (less coherent and localized). The fits agree qualitatively well with the data.}
    \label{fig:fits}
\end{figure*}

The numerical data for $\langle\sigma_z(t)\rangle$ obtained with inchworm diagrammatic Quantum Monte Carlo is fitted to the heuristic function given by Eq.~\eqref{eq:fit}. As illustrated in Fig.~\ref{fig:fits}, the fits match well the overall shape of the data across the different regimes, but in some cases, the error bars on the input data were scaled in order to ensure acceptable values of $\chi^2/$d.o.f. The nonlinear fit function has 7 parameters ($\Omega$, $\phi$, $\gamma_1$, $\gamma_2$, $a$, $b$, $c$). Although the initial condition requires $\langle\sigma_z(0)\rangle=a\cos(\phi)+b+c=1$, strictly enforcing this condition led to worse fits. Instead the error bar on the $t=0$ point was set to have very small (typically between $10^{-4}$ and $10^{-8}$) but nonzero. Because the nonlinear fit has many fit parameters, the $\chi^2$-function can have multiple minima and thus the fit result may depend on the choice of the initial parameter guess for the $\chi^2$-minimization. The overall error bars on the fit parameters are comprised of the propagated statistical errors of the numerical data, and the systematic error related the variability of the minimization procedure.

\section{Additional plots of fit parameters}
\begin{figure*}[bht]
    \centering
    \includegraphics[width=0.5\textwidth]{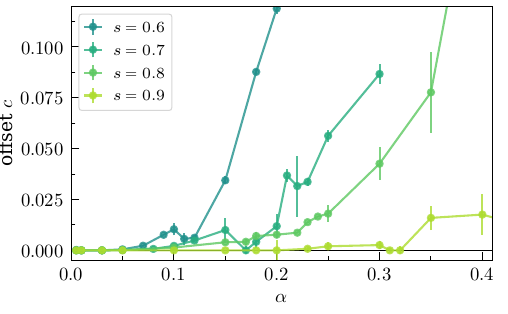}\hfill
    \includegraphics[width=0.5\textwidth]{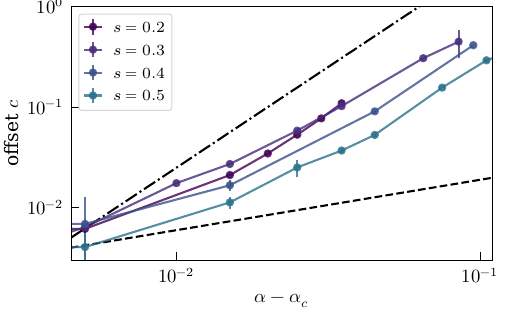}
    \caption{Left panel: the offset fit coefficient $c$ as a function of the coupling $\alpha$ for $s\geq0.6$ (data for $s\lesssim0.5$ is shown in the main text). While for small $s$ the dynamical localization transition is sharp, for larger values of $s$ the curves are smoother and extracting the transition points is more challenging. The right panel shows the offset values for $s\leq0.5$ on a log-log scale. The dashed line corresponds to the critical exponent of the equilibrium QPT ($\beta=1/2$) and the dot-dashed line corresponds to an exponent of 2, for comparison. It is clear from the plot that the corresponding scaling of the transient phase diagram is inconsistent with the equilibrium exponent.}
    \label{fig:fitcoeffs}
\end{figure*}

Figure~\ref{fig:fitcoeffs} shows a detailed analysis of the transient localization transition for all values of $s$ considered. For values $s\gtrsim0.5$ the offset curves are significantly less sharp than for smaller values of $s$ and extracting the transition points is more challenging, resulting in a wider confidence interval for the transient phase diagram shown in Fig.~4 of the main text. For the equilibrium QPT the critical exponent $\beta$ defined via $\langle\sigma_z\rangle\propto(\alpha-\alpha_c)^\beta$ equals $1/2$ for $s\leq0.5$. The dynamical data is inconsistent with the equilibrium behavior, indicating a scaling exponent above 1.

\begin{figure}
    \centering
    \includegraphics[width=0.6\textwidth]{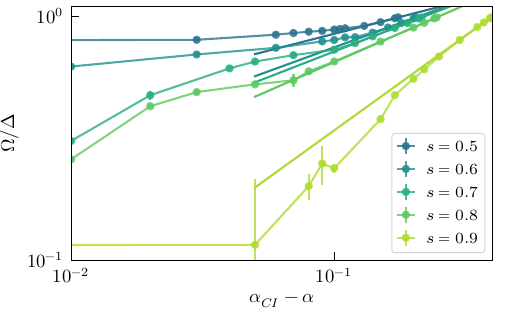}
    \caption{Oscillation frequency $\Omega$ as function of coupling $\alpha$ for $s\geq0.5$ on a log-log scale, which seems to indicate a power law behavior. The extracted powers range between 0.28 for $s=0.5$ and 0.78 for $s=0.9$.}
    \label{fig:omegascaling}
\end{figure}

Figure~\ref{fig:omegascaling} shows $\Omega$ plotted on a log-log scale, which indicates the existence of a power law.

\begin{figure*}
    \centering
    \includegraphics[width=0.5\textwidth]{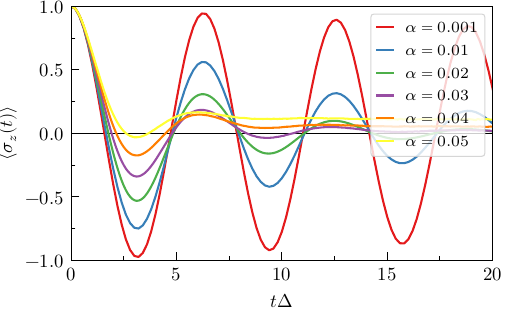}\hfill
    \includegraphics[width=0.5\textwidth]{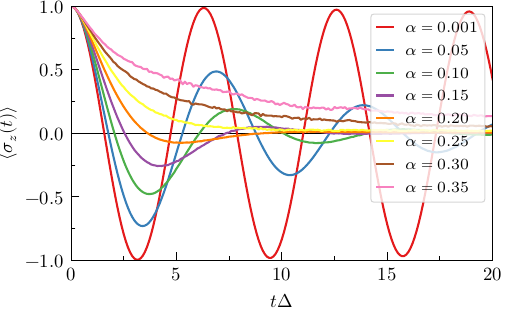}
    \caption{Time-evolution of $\langle\sigma_z(t)\rangle$ for $s=0.2$ (left) and $s=0.8$ (right) for select values of $\alpha$. This plot shows some of the data shown in Fig.~1 of the main text, but as a regular plot rather than a density plot. Not all $\alpha$ values shown in the main text are shown here, in order not to overcrowd the plots.}
    \label{fig:regulardynamics}
\end{figure*}

Figure~\ref{fig:regulardynamics} shows the time-evolution of $\langle\sigma_z(t)\rangle$ for $s=0.2$ and $s=0.8$, for some of the $\alpha$-values shown in Fig.~1 of the main text, but plotted as a set of curves, rather than combined into a density plot.
The density plot in the manuscript is generated from many such curves using linear interpolation.

\begin{figure*}
    \centering
    \includegraphics[width=0.5\textwidth]{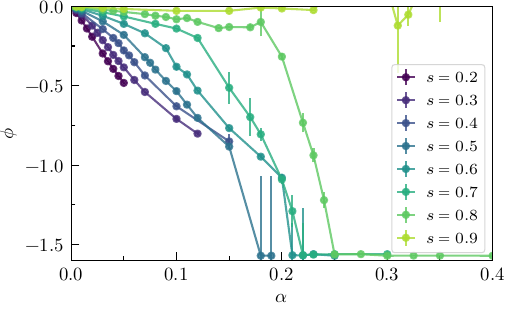}\hfill
    \includegraphics[width=0.5\textwidth]{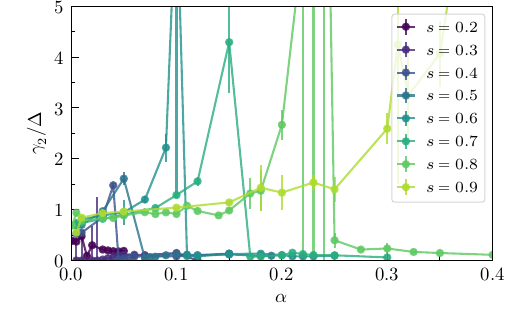}
    \includegraphics[width=0.5\textwidth]{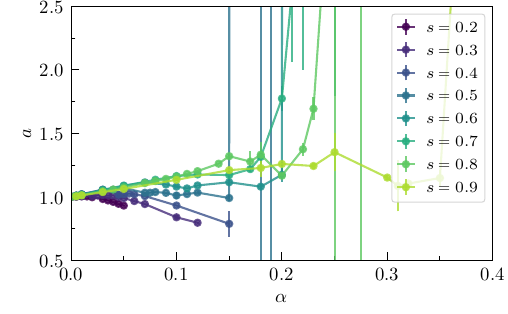}\hfill
    \includegraphics[width=0.5\textwidth]{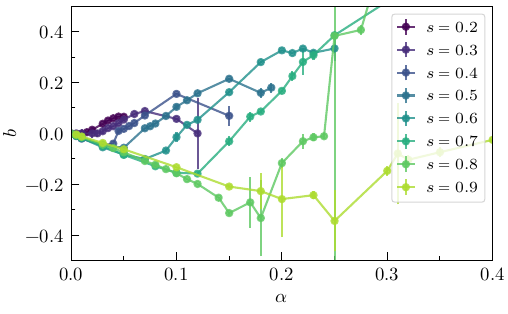}
    \caption{Additional fit parameters: phase $\phi$ (top left), overall damping $\gamma_2$ (top right), prefactor $a$ (bottom left), and prefactor $b$ (bottom right).}
    \label{fig:restcoeffs}
\end{figure*}

For completeness, Fig.~\ref{fig:restcoeffs} shows the remaining fit coefficients obtained by fitting our data to Eq.~\eqref{eq:fit}.

\end{document}